\documentclass[jap,preprint]{revtex4-1}
\usepackage{graphicx,float,amssymb,subfigure,dcolumn,epstopdf}
\usepackage{multirow}
\usepackage{bm}
\begin{document}
	 \title{Probing the Relationship between Anisotropic Magnetoresistance and Magnetization of ferromagnetic films}
	 \author{Wanli Zhang, Jing Chen, and Wenxu Zhang\footnote{Corresponding author. E-mail address: xwzhang@uestc.edu.cn}}
	 \affiliation{State Key Laboratory of Electronic Thin Films and Integrated Devices,
	 	University of Electronic Science and Technology of China, Chengdu, 610054, P. R. China}
	 \date{\today}
	 \begin{abstract}
      The anisotropic magnetoresistance (AMR) in thin permalloy strips was calculated at each steps during magnetization by the finite element method. The magnetization at equilibrium under different external fields was obtained by micromagnetic simulations, while the resistance with different magnetization was obtained by solving the Poisson equations iteratively until self-consistence. We find that the relation between magnetization and AMR deviates from the Stoner-Wohlfarth prediction when the magnetization is reduced from saturation. The reason is that the demagnetization is not necessarily from coherent rotation of the magnetic moment. We conclude that it is necessary to use numeric simulations to optimize the responses of AMR sensors.
	 \end{abstract}
	 \maketitle
	 \section{Introduction}
Since its discovery by William Thomson in 1857, the anisotropic magnetoresistance (AMR) effect, where the longitudinal resistance of ferromagnetic metals changes when the magnetic field is changed, has been one of the key methods to sense the magnetic field ranges from $10^{-5}$ Oe to several Oe \cite{mrbook}. Due to the process compatibility of silicon, it can be easily integrated onto silicon chips \cite{Popovic2006} so that the applications can be found in varies areas, including magnetic field sensing, current sensing and position or angle sensing. They are widely used in modern Internet of Things (IoT). 
 \par In order to design and optimize the performance of the sensing elements in real devices, besides the analytical expressions for Stoner-Wohlfarth (SW) models \cite{mrbook}, it is unavoidable to use numeric methods which can couple at least micromagnetics and Maxwell equations. Dynamic effect was investigated by using this strategy to model the transfer effect in magnetic heads \cite{Bruckner2013}. A multiscale modeling strategy based on phenomenological empirical equations is proposed to describe the performance of the magnetic field sensors \cite{Bartok2013}. It can be used to study the behavior of an AMR thin film sensor with material composition, crystallographic texture, and film thickness taken into account. The magnetoresistance change is found to increase significantly as the width of the wire is decreased \cite{Adeyeye1997}. Combination of micromagnetic modeling of magnetic moment distributions and the SW AMR model, exchange biased MR head was investigated by Koehler \emph{et al}. \cite{Koehler1993, Koehler1995}. In the SW model the change of the resistance $\Delta R/R$ is proportional to  $\cos^2\theta$ where $\theta$ is the angle between the magnetic moment and current density. Multiphysics program package \emph{Comsol} and micromagnetic simulations MuMax3 were used to calculate the AMR effect. The results of numerical simulations with allowance for inhomogeneous magnetization distribution showed that the proposed shape-coupled structures exhibit a significant increase in sensitivity as compared to that of classical barber-pole structures \cite{Dyuzhev2016}. 
 \par However, the simulations above calculate the AMR with the assumption that the current is parallel to the external field, therefore do not treat the current and the AMR in a self-consistent way: At fixed moment, the resistance and the current are independent upon each other. In their calculations, the resistance was fixed to the values corresponding to the current parallel to the magnetic field.  Self-consistent calculations were recently done by Fangohr \emph{et al}. \cite{Bordignon2007} and Abert \emph{et el.} \cite{Abert2013}, where the micromagnetic problem was coupled with the electrostatic problem and the two sets of equations were solved by the finite element method \cite{Bordignon2007}. The work shows that it is important to take into account the spatial variation of the current density when computing AMR.

\par Based on the SW model and general arguments of the symmetry, the AMR is proportional to $\cos^2\theta$, where $\theta$ is the angle between the magnetic moment of the samples and the current. However, in real materials, the magnetization are different in directions which is dependent on positions. Domain walls will generally appear when the sample dimensions exceed the single domain criteria. How will the AMR deviate from the $\cos^2\theta$ relations? The question can be answered by numeric simulations base on the micromagnetics. In this work, finite element based micromagnetic simulations were used to studied the magnetic reversal processes of nano element. Our simulations find that there is  are noticeable differences between the AMR calculated numerically and approximated by proportional to magnetization. This is the result of complex magnetic moment distribution. Regarding to this results, in order to optimization of the performances of AMR sensors, it is crucial to use numeric simulations rather than the simple coherent moment rotation models.
\section{Computation models and experimental details}
We use a nano thin plate elements with fixed length ($L$), width ($W$) and thickness ($t$). The element was meshed into tetrahedrons with the maximum edge length of 1.0 nm. Magnetic moment distributions at different magnetic fields were obtained with \emph{Nmag} which provides a systematic approach to multiphysics simulations with finite-element method \cite{nmag}. The dynamics of the magnetic moment under external field is governed by the Landau-Lifschitz-Gilbert (\emph{LLG}) equation:
\begin{equation}
\frac{d\vec M}{dt}=-\frac{\gamma}{1+\alpha^2}\left( \vec M\times \vec H_{eff}+\frac{\alpha}{M}\vec M\times(\vec M\times \vec H_{eff})\right),
\end{equation}
 where $\gamma$ is the gyromagnetic ratio, $\alpha$ is the damping coefficient and the auxiliary effective field $\vec H_{eff}$ includes the contributions from various effects of different nature:
 \begin{equation}
 \vec H_{eff}=\vec H_{ext}+\vec H_{anis}+\vec H_{exch}+\vec H_{demag}+\cdots.
 \end{equation}

\par
\begin{figure}
\includegraphics[scale=0.45]{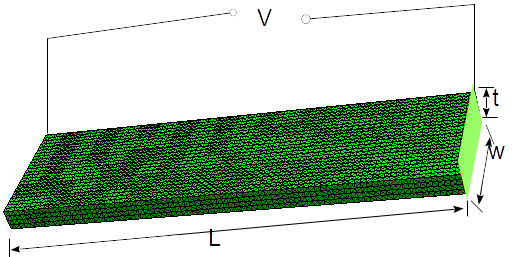}
\caption{\label{fig:mode} Schematic drawing of samples for micromagnetic electrostatic simulations, with voltage $V$ measured at the two faced surface of the orthogonal block with specified length ($L$), width ($W$) and thickness ($t$). }
\end{figure}

 The problem of finding the resistance (R) of the sample can be solved by using the Poisson equations
\begin{equation}
\nabla(\widehat \sigma(\vec r,\vec m) \nabla U(\vec r))=0, \label{equ:poison}
\end{equation}
 where $\widehat\sigma(\vec r,\vec m)$ is the position dependent electron conductivity matrix depending on the local magnetic moment ($\vec m(\vec r)$), and $U(\vec r)$ is the electric potentials. In more details, the electric field $\vec E$ is computed via
 \begin{equation}
 \vec E=-\nabla U(\vec r),
 \end{equation}
 and the current density
 \begin{equation}
 \vec J (\vec r)=\widehat \sigma (\vec r) \vec E=-\widehat \sigma(\vec r)\nabla U(\vec r)
 \end{equation}
  with suitable boundary conditions. In the electrostatic problem, the divergence of the current density must vanish
 \begin{equation}
 \nabla \cdot \vec J(\vec r)=0.
 \end{equation}
  When combined Equ.(4),(5) and (6), the Poisson equation for the potential is obtained as in Equ. (\ref{equ:poison}).
 Using the same mesh as in micromagnetics, Equ. (\ref{equ:poison}) was solved by \emph{FEniCS} \cite{fenics}. The resistivity $\rho=\sigma^{-1}$ at different vertexes were set to be $\rho(\theta)=\rho_0(1+\alpha\cos^2\theta)$, where $\rho_0$ is the resistivity of the material in the absence of AMR and $\alpha$ is the AMR ratio. In our simulations, we used $\sigma_0=5.99\times 10^7 S/m$ and $\alpha=0.5\%$. These two parameters do not influence the discussions of the AMR effects in our work due to the fact that the voltage and its variation scale with them. The Poisson equation was solved iteratively until the current density distributions converge to $10^{-8}$. The resistance between two electrodes was calculated by $R=V/I$ where $I$ is the integration of current normal to the electrodes, and $V$ is the voltages set between the electrodes as the boundary condition for the problem.
 A schematic drawing of our samples with the FEM net are shown in Fig. \ref{fig:mode}. The geometric parameters are set to $L=210$ nm, $W=60$ nm and $t=10$ nm. The electrodes are defined at the opposite sides of the sample.
\section{Results and discussions}
\begin{figure}
\centering
 \includegraphics[width=0.45\textwidth]{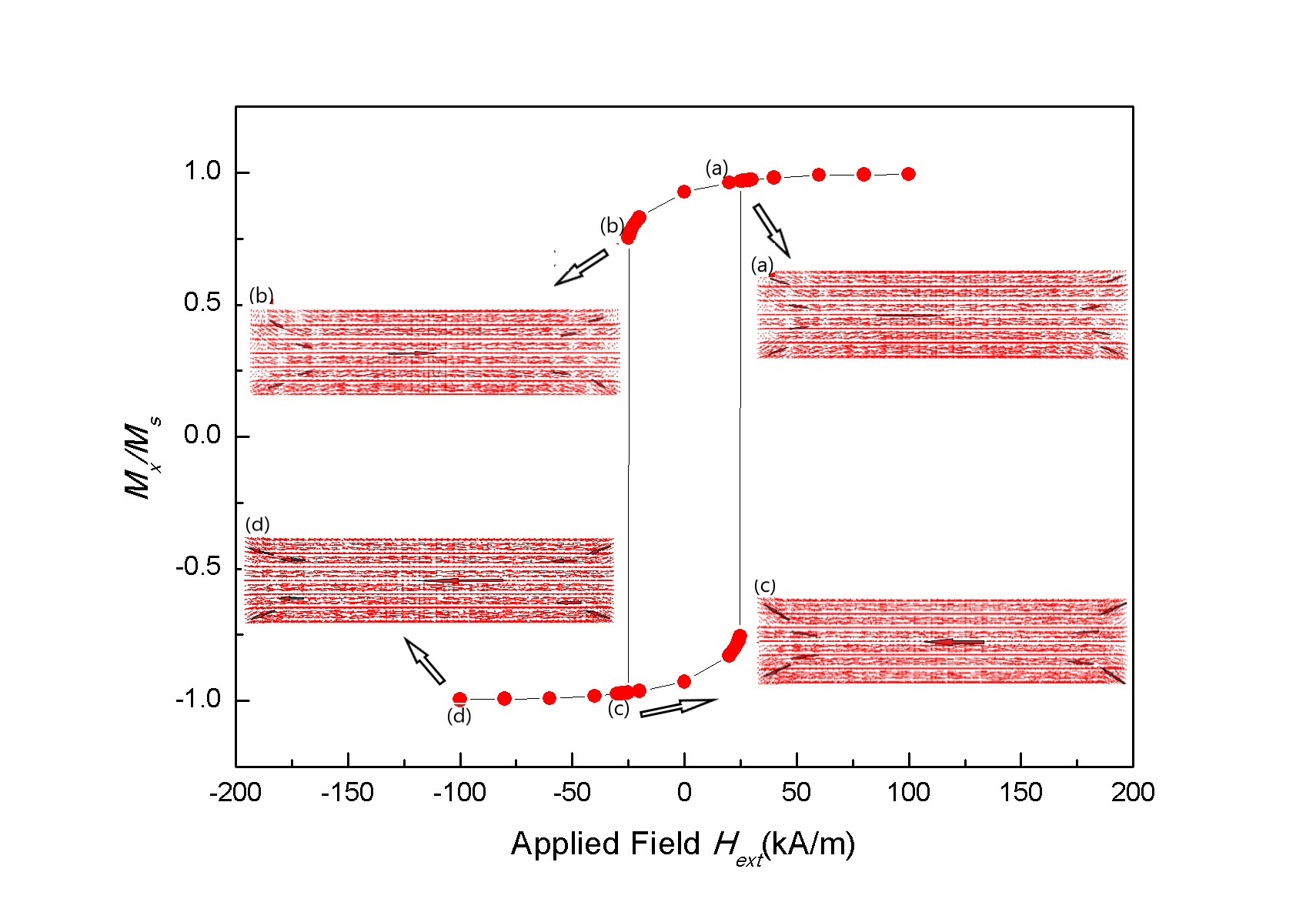}
\caption{\label{fig:mom} The hysteresis loop of the rectangle element used in this simulation and the magnetic moment distributions at different external fields (a) 23.0 kA/m, (b)-24.6 kA/m,(c)-25.0 kA/m and (d) -100 kA/m.}
\end{figure}

For the typical square plate shaped elements, the demagnetization field  will cause a small misaligned moment at both ends forming a ``flower''-state when the magnetic field is decreased from the initial positive one as shown in Fig.\ref{fig:mom} where the hysteresis loop is obtained by micromagnetic simulations. Decrease the field leads to the gradual reversal of the moment as required to minimize the free energy of the systems as from (a) to (b). The reversal regions expands. When the field reaches the critical value, here, is about 25 kA/m antiparallel to the initial field, a first order transition of the magnetic state occurs. This leads to the flip of the magnetic moment to the negative value as in (b) and (c). The magnetization rapidly saturates in the negative direction and in (d), when the negative field increases. Then the field is decreased and then reversed.

\begin{figure}
\includegraphics[scale=0.2]{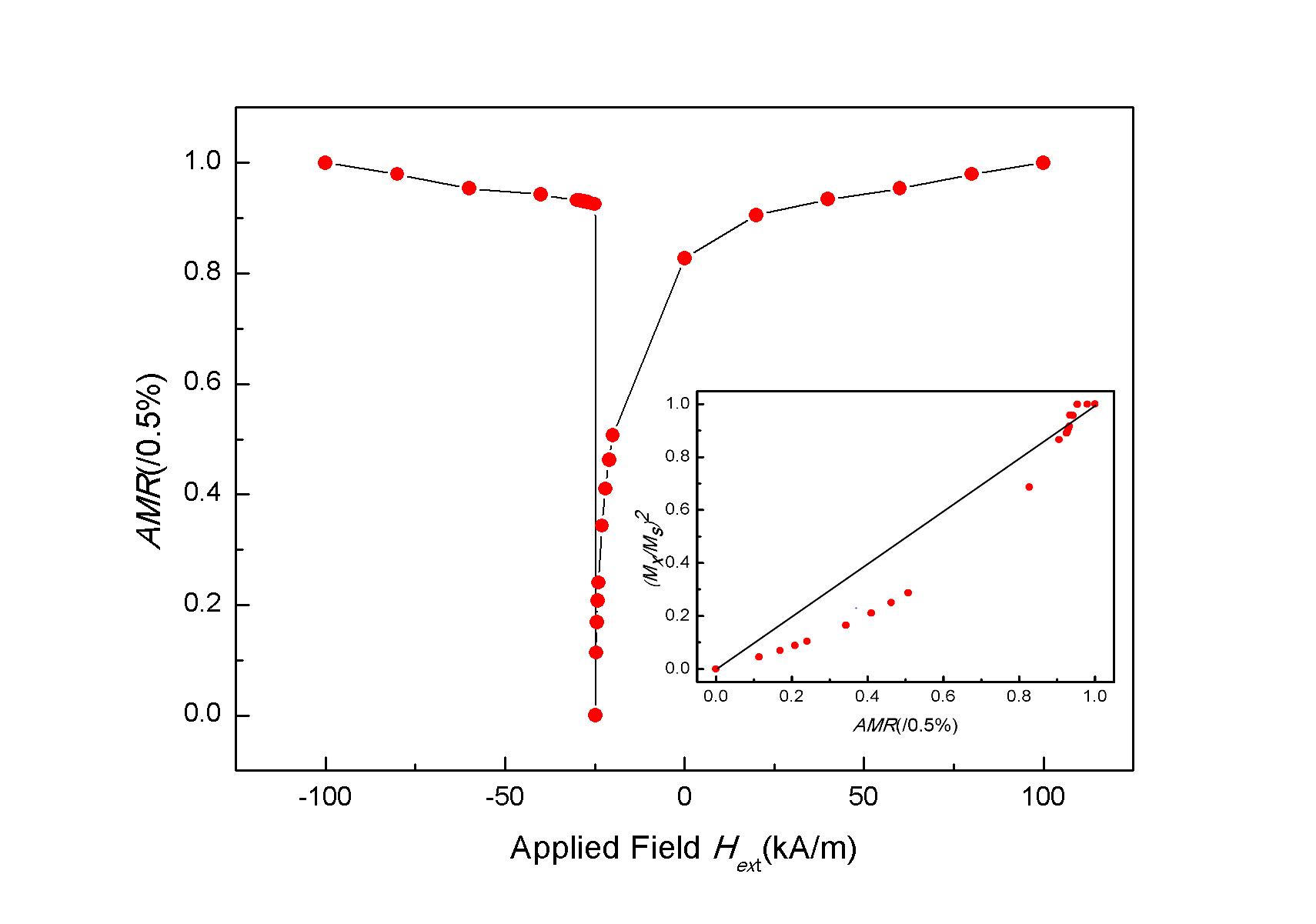}
\caption{\label{fig:amr-loop} The calculated AMR of the rectangle element. Relation between the AMR and $\left<(\frac{M_x}{M_s})^2\right >$ is shown in the inset. The data are shown by the filled circles, while the solid lines are guide for eyes.}
\end{figure}

\par
\begin{figure}
\includegraphics[scale=0.25]{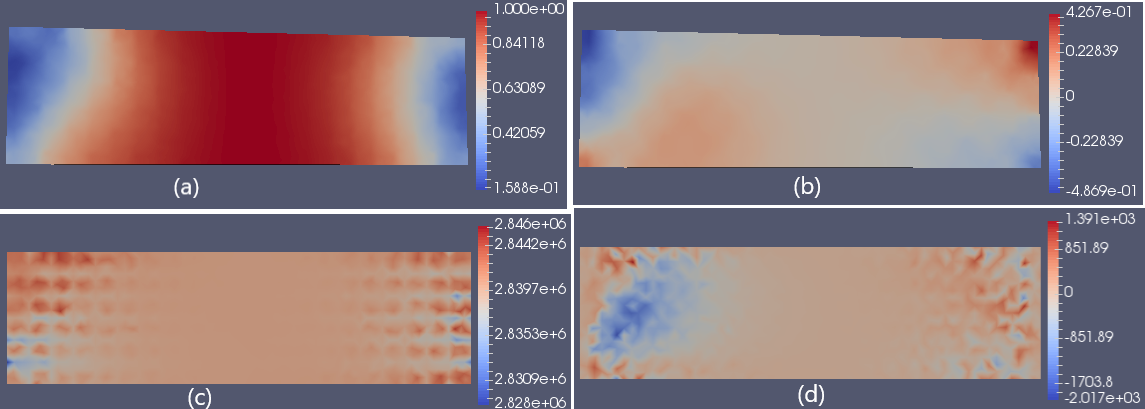}
\caption{\label{fig:jm} The position dependent magnetic moment $m_x$(a), $m_y$(b) and the current density $j_x$(c) $j_y$(d) in the x-y plane of the sample.}
\end{figure}

 During this magnetic field variation, the AMR of the sample calculated are shown in Fig. \ref{fig:amr-loop}. We can see that during the process of when the magnetization decrease, the AMR decreases with it. It reaches the minimum at the critical point and shoots up when the magnetic moment switched to the opposite direction. In order to investigate the relationship between the AMR and variation of magnetization, we plot the $\left<(\frac{M_x}{M_s})^2\right >$ , where $\left<\cdot \right>$ is the average over the finite element cells, as the function of AMR shown in the inset of Fig. \ref{fig:amr-loop}. In the ideal SW model they should be on a straight linear because of the relationship of $AMR\propto\left < \cos^2\theta \right >$ defined above. In the initial stage of decrease of the moment, when the magnetic moment is more than 80\% of the initial values, the linearity is reasonable. However, the deviation from linearity is obvious as in the inset when the magnetic moment is further reduced. It gives a positive contribution to the AMR value, which can be as large as 50\%. The reason for this nonlinearity comes from the noncollinearity of the magnetic moment distribution, which causes noncollinearity of the current density in the cells with respect to the magnetic moment. The distributions of the current density are shown in Fig. \ref{fig:jm}. In the region where the magnetic moment is deviated from the magnetic field, the resistance on the path through which the current pass is also different from the homogenous region. However, the distribution of the current density does not take a copy of the magnetic moment. At the same time, when the decrease of the magnetization does not lead to a change of the magnetic moment direction with respect to the current density, they do not contribute to the AMR effect. For example, during the Bloch wall displacement, the magnetization varies while the AMR retains. In this case, there is no hope to get a linear response of the magnetoresistance with respect to the square of the magnetization. With regarding to these results, it is crucial to form a homogenous magnetized samples to achieved good linearity in the magnetic sensors. At the same time, it is necessary to use numeric simulations to optimize the performances of magnetic field sensor, so that the linearity can be optimized by proper design of barber poles.
%\par
%The nonlinearity is also confirmed by experiments. The AMR and magnetization are both The relationship of magnetization and AMR values is shown in Fig.\ref{fig:amr-m_exp}. The hysteresis loop and AMR of thin films are displayed in the insets of the figure. XXX The deviation is large at the stage... For this thin films, the magnetization reversal is realized in different modes as well studied in the literatures. At the initial stage, magnetic domains and domain walls are formed, which will cause a large portion of nonlinearity.

\section{Conclusions}
To summarize, we have shown in this work, the magnetization and the AMR are not proportional as suggested by the coherent moment rotation model. The proportionality is only good when the magnetization variation is less than 80\% of its saturation value. It is deviated from the analytical model prediction when the magnetic moment is far from saturation.  We find that it is due to the formation of inhomogeneous magnetization region. Our work points to the conclusion that the joint self-consistent solution of LLG equations in micromagnetics and Poisson equations in electrostatics is necessary to understand and optimization the performances of AMR sensors.
\section{Acknowledgments}
Discussions with Fangohr were acknowledged. This work is supported by National Key R\&D Program of China (No.2017YFB0406403 ) from MOST.

%\bibliographystyle{unsrt}
%\bibliography{ref}

\begin{thebibliography}{10}

\bibitem{mrbook}
S.~Tumanski.
\newblock {\em Thin Film Magnetoresistive Sensor}.
\newblock IOP Publishing Ltd, 2001.

\bibitem{Popovic2006}
Radivoje~S. Popovic, Predrag~M. Drljaca, and Pavel Kejik.
\newblock {CMOS} magnetic sensors with integrated ferromagnetic parts.
\newblock {\em Sensors and Actuators A: Physical}, 129(1-2):94--99, may 2006.

\bibitem{Bruckner2013}
Florian Bruckner, Christoph Vogler, Bernhard Bergmair, Thomas Huber, Markus
  Fuger, Dieter Suess, Michael Feischl, Thomas Fuehrer, Marcus Page, and Dirk
  Praetorius.
\newblock Combining micromagnetism and magnetostatic maxwell equations for
  multiscale magnetic simulations.
\newblock {\em Journal of Magnetism and Magnetic Materials}, 343:163--168, oct
  2013.

\bibitem{Bartok2013}
Andr{\'{a}}s Bart{\'{o}}k, Laurent Daniel, and Adel Razek.
\newblock A multiscale model for thin film {AMR} sensors.
\newblock {\em Journal of Magnetism and Magnetic Materials}, 326:116--122, jan
  2013.

\bibitem{Adeyeye1997}
A.~O. Adeyeye, J.~A.~C. Bland, C.~Daboo, D.~G. Hasko, and H.~Ahmed.
\newblock Optimized process for the fabrication of mesoscopic magnetic
  structures.
\newblock {\em Journal of Applied Physics}, 82(1):469--473, jul 1997.

\bibitem{Koehler1993}
T.~R. Koehler, Bo~Yang, Wenjie Chen, and D.~R. Fredkin.
\newblock Simulation of magnetoresistive response in a small permalloy strip.
\newblock {\em Journal of Applied Physics}, 73(10):6504--6506, may 1993.

\bibitem{Koehler1995}
T.R. Koehler and M.L. Williams.
\newblock Micromagnetic modeling of a single element {MR} head.
\newblock {\em {IEEE} Transactions on Magnetics}, 31(6):2639--2641, 1995.

\bibitem{Dyuzhev2016}
N.~A. Dyuzhev, A.~S. Yurov, R.~Yu. Preobrazhenskii, N.~S. Mazurkin, and M.~Yu.
  Chinenkov.
\newblock Shape-coupled magnetoresistive structures: a new approach to higher
  sensitivity.
\newblock {\em Technical Physics Letters}, 42(5):546--549, may 2016.

\bibitem{Bordignon2007}
Giuliano Bordignon, Thomas Fischbacher, Matteo Franchin, Jurgen~P. Zimmermann,
  Alexander~A. Zhukov, Vitali~V. Metlushko, Peter A.~J. de~Groot, and Hans
  Fangohr.
\newblock Analysis of magnetoresistance in arrays of connected nano-rings.
\newblock {\em {IEEE} Transactions on Magnetics}, 43(6):2881--2883, jun 2007.

\bibitem{Abert2013}
Claas Abert, Lukas Exl, Florian Bruckner, Andr{\'{e}} Drews, and Dieter Suess.
\newblock magnum.fe: A micromagnetic finite-element simulation code based on
  fenics.
\newblock {\em Journal of Magnetism and Magnetic Materials}, 345:29--35, nov
  2013.

\bibitem{nmag}
Thomas Fischbacher, Matteo Franchin, Giuliano Bordignon, and Hans Fangohr.
\newblock A systematic approach to multiphysics extensions of
  finite-element-based micromagnetic simulations: Nmag.
\newblock {\em {IEEE} Transactions on Magnetics}, 43(6):2896--2898, jun 2007.

\bibitem{fenics}
Anders Logg, Kent-Andre Mardal, and Garth Wells, editors.
\newblock {\em Automated Solution of Differential Equations by the Finite
  Element Method}.
\newblock Springer Berlin Heidelberg, 2012.

\end{thebibliography}

%\end{thebibliography}

\end{document}